\documentstyle[preprint,aps]{revtex}
\begin{document}
\draft

\title{Free Thermal Convection Driven by Nonlocal Effects}

\author{Jorge Ibsen, Rodrigo Soto and Patricio Cordero
\\ Departamento de F\'{\i}sica, Facultad de Ciencias F\'{\i}sicas y
Matem\'aticas \\ Universidad de Chile, Casilla 487, Santiago 3, Chile \\
e-mail: pcordero@tamarugo.cec.uchile.cl}

\maketitle

\begin{abstract}
We report and explain a convective phenomenon observed in molecular dynamics
simulations that cannot be classified either as a hydrodynamics instability
nor as a macroscopically forced convection. Two complementary arguments show
that the velocity field by a thermalizing wall is proportional to the ratio
between the heat flux and the pressure. This prediction is quantitatively
corroborated by our simulations.
\end{abstract}


Free thermal convection ---driven by buoyancy or by surface tension--- is a
perfectly well understood phenomenon derivable from Navier Stokes equations
\cite{Chandr,NPV}.  Simulations of free thermal convection by means of {\em
Molecular Dynamics} (MD) techniques can be achieved with systems with as few as
$10^3$ particles and already  these small systems exhibit hydrodynamic
behavior as seen for example in \cite{Mare87}-\cite{RC93}.  Moreover MD is
useful to study fluid  phenomena at the microscopic level without having to
make assumptions concealed behind  the Navier-Stokes equations such as the
Fourier law, Newton's law of viscosity and local thermodynamic equilibrium.

In the following we are going to present a unusual convective phenomenon (not
predictable by Navier-Stokes equations) related to the variation of the
temperature field in one mean free path $\ell$ through the adimensional
parameter $\ell\nabla T/T$. This parameter can be interpreted as a measure of
how far from local thermal equilibrium the system is at a given point. When
effects violating the local thermodynamic equilibrium are present one should
question the very concept of temperature but we will manage without engaging
in such delicate matters.

The convective phenomenon that we are reporting takes place when there is a
temperature gradient parallel to a thermal wall. The mechanism can be sketched
as follows: the particles that approach a point $P$ of the wall come from an
anisotropic distribution while the particles that hit the wall at $P$ come
back to the system with a distribution which is isotropic, or at least less
anisotropic than the incoming flux. A careful assessment of the difference
between the incoming and outgoing fluxes at $P$ yields the conclusion that
there is a net mass flux parallel to the wall.  We have observed this
phenomenon in MD simulations and have made a theoretical estimation of its
value. In real experiments the effect will be small but it should be
observable in a rarefied gas.

We have made MD simulations of a two dimensional gas of hard disks in a square
box, using our own efficient algorithm\cite{MRC93} and the carefully devised
measurement routines described in \cite{Dino}.  Each numerical experience
consisted of two runs: \ {\bf (1)}~The system with periodic vertical walls was
subjected to a temperature difference, relaxed for 200 thermal diffusion times
$t_T$, and then the temperature profile $T(y)$ was carefully measured for
another 200 $t_T$. \ {\bf (2)}~A second simulation was run under the same
conditions as in (1), except that a periodic {\em permeable thermal vertical
wall} was added. This new wall was defined to have the previously obtained
profile $T(y)$, namely each particle hitting the vertical wall emerged on the
other side of the box with a velocity taken from a heat bath at the local
temperature $T(y)$. The sign of the vertical component of this velocity was
random, and therefore microscopically it is a non slip boundary condition, in
the sense that the emerging particles do not remember the velocity with which
they came. Letting the particles pass through the (periodic) wall is totally
irrelevant to the resulting phenomenon, but it helps reducing the boundary
effects near the wall. Again the system was relaxed for 200 $t_T$, and then
measurements were averaged in time during the next $600
\, t_T$. The measurements where done dividing the system in square cells.
Densities and the velocity field were measured in every cell, and fluxes were
measured across the cell walls.

Units are chosen such that particle's mass and diameter, the Boltzmann
constant, and the temperature at the bottom, $m$, $D$, $k_{B}$ and $T_{b}$
respectively, are fixed to unity. With this particular choice of units the
lengths are in diameter units, the temperature in energy units, and the time
in units of $\sqrt{mD^2/k_{B}T_{b}}$. The control parameters of each
simulation are the number of particles $N$, the bulk number density $n_{B}$,
and the temperature at the top $T_{t}$ where $T_{t}\!<\!T_{b}$.

Our main simulation considered a system of $N\!=\!1444$ hard disks, bulk
number density $n_{B}\!=\!0.05$, implying a box side of 170 and a mean free
path of about 7, and at the top the temperature was fixed to be $T_t\!=\!0.1$.

The main observation is the following: a convective current stabilizes in the
neighborhood of the vertical wall, moving towards the warmer zone. In
figures~\ref{FIG-Veloc} and~\ref{FIG-Mass} it is possible to see the velocity
field $\vec v$ and the mass flux $ m n \vec v$.  At the bottom the convective
current necessarily bends towards the center to come  up along the central
part of the box. Since the gas is highly compressible, the eye of the
convective rolls are far from the expanded hotter zone. The velocity component
$v_{y}$ in the cells by the vertical wall is almost constant, and its average
was
\begin{equation}
v_y = -0.015 \pm 0.002  \qquad\hbox{[observed]}  \label{vy-experim}
\end{equation}
after excluding 10 cells in the upper and lower extremes, with $76\times 76$
the total number of cells.

The same convective phenomenon was observed in all the other situations we
simulated: \ (i) $n=0.01$, $N=8100$, $T_b=1.0$, and $T_t=0.01$, \ (ii)
$n=0.25$, $N=1444$, $T_b=1.0$, and $T_t=0.1$. The velocity component $v_{y}$
measured near  the vertical walls in (i)  was $v_{y} = 0.014 \pm 0.003$ and it
shows the same behavior as the preceding simulation. The vertical component of
the velocity in (ii) however is not longer constant, it increases with height.
The theoretical derivations that we make below are not applicable to this
denser case but it is interesting to observe that the phenomenon still exists.

Finally we made another simulation in which the temperature profile of the
thermalizing wall was not the one obtained from a first run but rather $T(y)$
was chosen arbitrarily to be a smooth  monotonic profile. In this case we used
$n=0.05$, $N=1444$ and $T_t=0.1$. Again a convective current was created with
similar characteristics to the previous ones. This result and the theoretical
calculations below suggest that to obtain this convective motion, it is enough
to have  a temperature gradient parallel to a thermalizing wall so that each
particle emerging from the wall comes from a (at least partially) thermalized
distribution.

This kind of convective motion can not be obtained within the usual frame of
Navier-Stokes hydrodynamics, unless these equations are solved imposing by
hand a nonvanishing tangential velocity as boundary condition at the vertical
wall.  That however would be artificial because hydrodynamics is built using
only the first three momenta of the distribution function which are not enough
information to take into consideration the phenomena that we are reporting.

In what follows, we give two heuristic and complementary derivations for a
rarefied two dimensional hard disk gas, one based on local nonequilibrium
distribution functions, and the other one based on the mean free path theory of
transport.  Both derivations yield essentially the same prediction for the
velocity field near the vertical thermalizing wall. What the calculations
below imply is that the vertical wall exerts an effective tangential force on
the gas such that a velocity field --- proportional to the ratio between the
heat flux and the pressure --- pointing against the heat flux is established.

The basic idea behind the following two derivations is that particles hitting
the thermalizing wall at a point $P$ (see figure~\ref{FIG-Wall}) come from an
anisotropic nonequilibrium environment, while the particles emerging from $P$
come from an equilibrium isotropic distribution. It is understandable then
that some fluxes do not necessarily cancel and in particular we prove that
there is a net velocity field.

\bigskip

\noindent {\sc Nonequilibrium Interpretation:} \ The velocity
distribution function near a point $y$ of the thermalizing wall
(figure~\ref{FIG-Wall}) has two contributions:  (a) one from the  particles
that come towards $y$ from a nonequilibrium velocity distribution and (b) the
other one from the outgoing particles that come from the thermal bath at $y$.
The nonequilibrium distribution function for a system under a heat flux
adapted from \cite{Grad} to the case of a two dimensional system is
\begin{equation}
f_{\hbox{neq}} = \left(1+\frac{m}{2pT}\left[
\frac{mv^{2}}{2T}-2\right] \, \vec{v} \cdot\vec{q}\right) \, f_{\hbox{eq}}
\end{equation}
where $f_{\hbox{eq}}$ is the usual Maxwellian distribution. Then the velocity
distribution near the right wall is %
\begin{equation}
f = \left\{ \begin{array}{cc} f_{\hbox{neq}}(\vec{v}) & v_{x} > 0 \\
f_{\hbox{eq}}(\vec{v}) & v_{x} < 0 \end{array} \right.
\end{equation}

Using this distribution, the $x$-$y$ component of the stress exerted by the
fluid against the wall ($\sigma_{xy} = mn\langle v_x v_y
\rangle_{f}$) is
\begin{equation}
\sigma_{xy} = \frac{q}{8} \sqrt{\frac{2m}{\pi T}}
\end{equation}

Then the force per unit length exerted by the wall on the fluid is, by the
action-reaction principle, the negative of the previous expression, hence it
points {\em antiparallel to the heat flux}.

The velocity near the wall can be estimated using Newton's law ($\sigma_{xy}=
\eta\partial v_{y}/\partial x$ where $\eta$ is the shear viscosity) and the
assumption that this velocity decays at distances comparable with the mean
free path $\ell$  and it is,
\begin{equation}
v_{y} = -\frac{\ell \sigma_{xy}}{\eta}
\end{equation}

Replacing the expressions for the shear viscosity, the mean free
path, and using the equation of state for an ideal gas yields ~\cite{McQ}.
\begin{equation}
 v_{y} = -\frac{1}{8} \frac{q}{p} \label{VY}
\end{equation}

\bigskip

\noindent {\sc Kinetic Interpretation:} Let us consider the mass flux balance
at an arbitrary point $P$ of the wall, as shown in figure~\ref{FIG-Wall}. The
mass flux coming from an angle between $\phi$ and $\phi + d\phi$ with respect
to the normal to the wall and reaching $P$ is %
\begin{equation}
d\vec{\jmath} =  m\, n \sqrt{\frac{ T}{m}}\,a(\phi) (\cos\phi,-\sin\phi)\,
d\phi
\end{equation}
 where $n$ and $T$ are the number density and temperature at the points where
the particles come from whereas $a(\phi)$ is a geometrical factor
depending on the incident angle. We do not give a value for $a(\phi)$ since it
is well known that the mean free path theory of transport is too simple to
produce the correct numerical factor~\cite{McQ}. Since the number density is
small then $p=n\,T$.

The combined inward mass flux from the directions $\phi$ and $-\phi$ to $P$ is
then
\[
d\vec{\jmath}_{\hbox{in}}(\phi) = {\sqrt{m}\,p}\, a(\phi)\left[
\cos\phi \left( \frac{1}{\sqrt{T(\phi)}} + \frac{1}{\sqrt{T(-\phi)}} \right)
\, \hat{x} -\sin\phi \left( \frac{1}{\sqrt{T(\phi)}} -
\frac{1}{\sqrt{T(-\phi)}} \right) \, \hat{y} \right] d\phi
\]

The combined outward mass flux from directions $\phi$ and $-\phi$ can only be
in the $\hat x$ direction, since the flux  comes from the local equilibrium
distribution at the wall, and it is $d\vec{\jmath }_{\hbox{out}}(\phi) =-
m\,n\,\sqrt{\frac{T}{2 \pi m}}\cos\phi \, \hat{x}\, d\phi $. The density $n$
in $d\vec{\jmath}_{\hbox{out}}$ can not be replaced by $p/T$, because the
equation of state is not valid at wall points.  This number density is
unknown and can be determined imposing  null net mass flux in the $\hat x$
direction.

The total flux in the $\hat y$ direction is %
\begin{equation}
\vec{J}\cdot\hat{y} = \int_{0}^{\pi/2}
(d\vec{\jmath}_{\hbox{in}}+d\vec{\jmath}_{\hbox{out}})\cdot\hat{y}
 = A\sqrt{m}\,\frac{p\,\ell}{T^{3/2}}\,\frac{dT}{dy}
\end{equation}
with
\[A=\int_{0}^{\pi/2} a(\phi)\sin^{2}\phi d\phi \]
and where we have used that $T(\phi)= T + \ell\sin\phi\, dT/dy$ with $\ell$
the mean free path. It must be remarked that the previous expression can be
written as %
\begin{equation}
\vec{J}\cdot\hat{y} = const \times n v_{\hbox{th}} \left(
\frac{\ell}{T}\frac{dT}{dy}\right)
\end{equation}
where $v_{\hbox{th}}$ is the thermal velocity. The previous result implies
that a mass flux parallel to the temperature gradient is induced near the
wall, and it is proportional to the adimensional parameter $\frac{\ell}{T}
\frac{dT}{dy}$, which is a measure of how far from local thermal equilibrium
the system is at a given point.

The velocity field near the wall can be estimated dividing this flux by the
mass density, using the Fourier law, the expressions for the mean free path,
the thermal conductivity, and the equation of state for an ideal
gas~\cite{McQ}.  The outcome is %
\begin{equation}
v_y = -\frac{A}{4}\sqrt{\frac{\pi}{2}} \frac{q}{p}
\end{equation}

This result predicts the same behavior as~(\ref{VY}).

\medskip

{}From these two heuristic arguments we have shown that the mechanism for this
type of convection is already present in nonequilibrium thermodynamics and in
fact both formulations, mean free path theory  and nonequilibrium local
distribution functions, give essentially the same result: \ the velocity field
near the wall is proportional to the heat flux, and it is independent of the
point $P$ ---due to the absence of external forces and energy sources, $p$ and
$q$ are uniform--- as  it can be  appreciated in figure~\ref{FIG-Veloc}. Our
predicted value for $v_y$  from our observations of $q$ and $p$ is %
\begin{equation}
 v_{y} = -0.016 \pm 0.003 \quad \hbox{[predicted]}
\end{equation}
which should be compared with (\ref{vy-experim}), and $v_{y} = -0.020 \pm
0.001$ for the simulation with $n=0.01$, $N=8100$, $T_b=1.0$, and $T_t=0.01$.

The extension to three dimensions is straightforward giving essentially the
same result, indicating that this convective motion could be observed in a
rarefied gas. For example using gaseous Helium at atmospheric pressure  with a
temperature gradient of $\nabla T= 100 [K/cm]$, the velocity near the wall
that we are predicting is  $v_y=3.8 [mm/s]$, but it may be somewhat smaller
because, in a real experiment, the average flux coming out from every point at
the thermalizing wall is not totally isotropic.

Regarding the heat flux $\vec q$, there has been an interesting recent
proposal applicable to the case of systems under a heat flux, saying that one
should observe a departure from the standard Fourier law \cite{CVzz} (which
actually motivated our series of simulations.) \  What we observe (see
figure~\ref{FIG-Energy}) is that the energy flux is consistent with the
Fourier law plus the kinetic flux $\frac{1}{2}\langle m\,n\,v^2\,\vec v\,
\rangle$ of an ideal gas.  The effect predicted in \cite{CVzz} for our system
is about an order of magnitude smaller that the total flux, and since fluxes
are noisier than densities, we can not yet see if such an effect exists.

In summary, we have observed and justified that there exists a free thermal
convection, which stems from non local effects due to the presence of
nonequilibrium distributions. The velocity by the wall is proportional to
$\ell \nabla T/T$, which is assumed to vanish in hydrodynamics (local
thermodynamic equilibrium).

It is a pleasure to thank Rosa Ram\'{\i}rez and
Dino Risso for fruitful discussions. This research has been supported by the
{\em Fondecyt} grant 193 1105. \ J.I. would like to thank {\em Fundaci\'on
Andes} finantial support through grant C-12400 and \ R.S. a {\em Conicyt}
Doctoral Fellowship and the Universidad de Chile Postgraduate Scholarship
PG-126-94.

\newpage

\begin{figure}
\caption{Velocity field measured using MD simulations. The horizontal walls
are kept at uniform temperature, the warmer wall is at the bottom, and the
vertical walls have a different temperature at each point. The number of
particles, the bulk number density, and the top temperature were
$N=1444$, $n=0.05$, and $T_{t} =0.1$ respectively.}
\label{FIG-Veloc}
\end{figure}

\begin{figure}
\caption{Mass flux measured in the same simulation shown in the previous
figure.}
\label{FIG-Mass}
\end{figure}

\begin{figure}
\caption{Contributions to the mass flux from different directions. Particles
come from regions at different temperatures and emerge with velocities from an
isotropic distribution at temperature $T$.}
\label{FIG-Wall}
\end{figure}

\begin{figure}
\caption{Energy flux measured in the same simulation as in previous
figures. The $x$ component has been amplified four times to show how it is
distorted by the convective current.}
\label{FIG-Energy}
\end{figure}

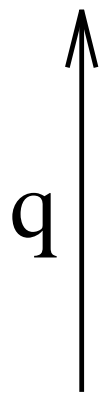
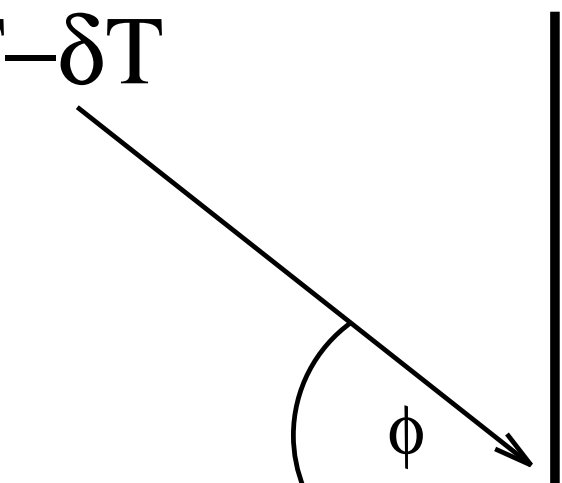
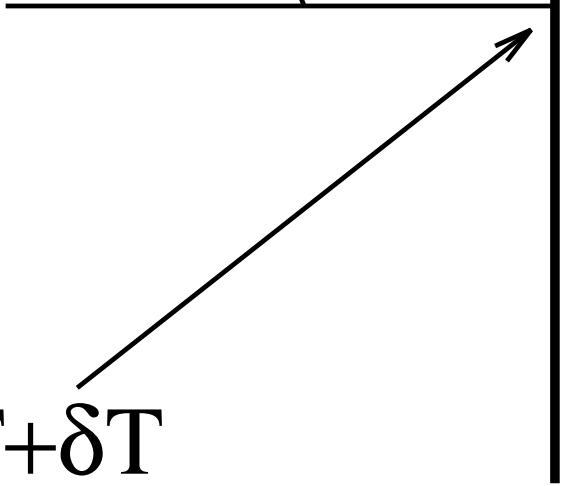
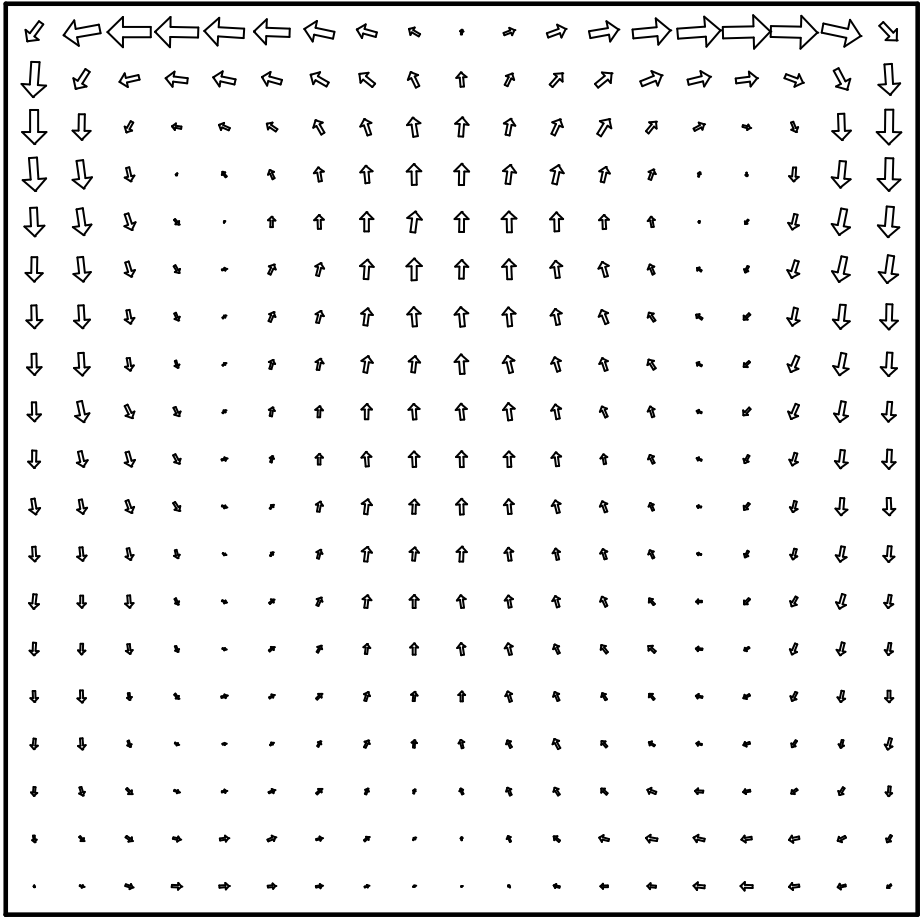
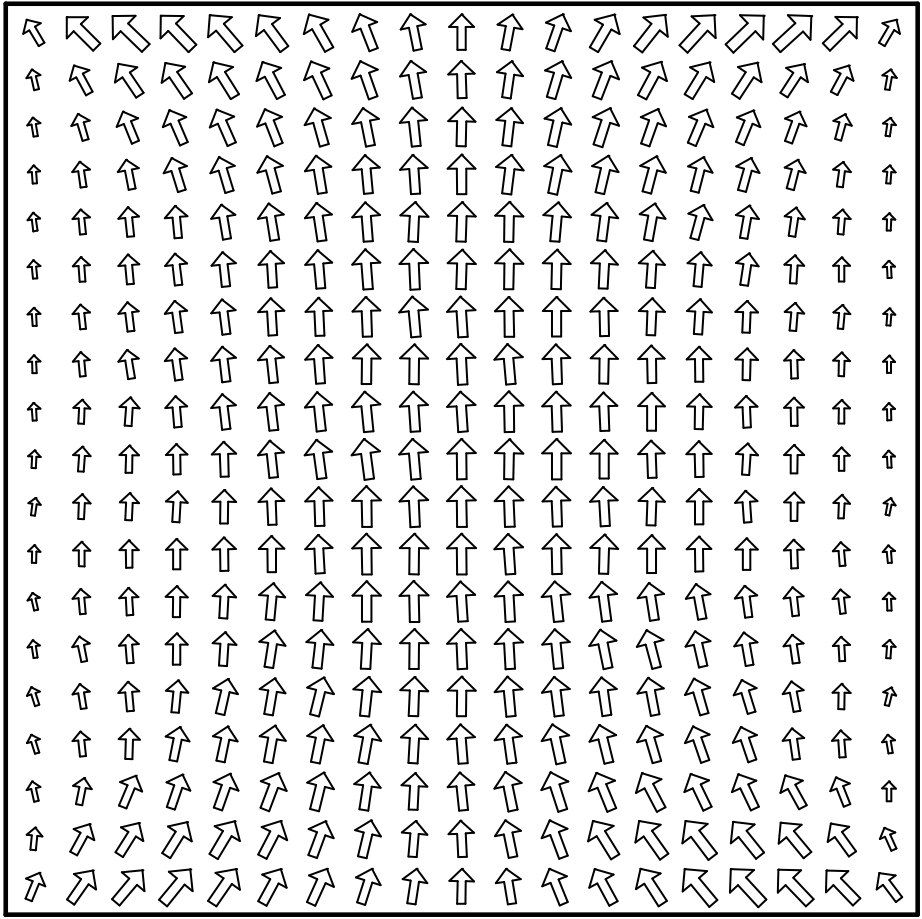
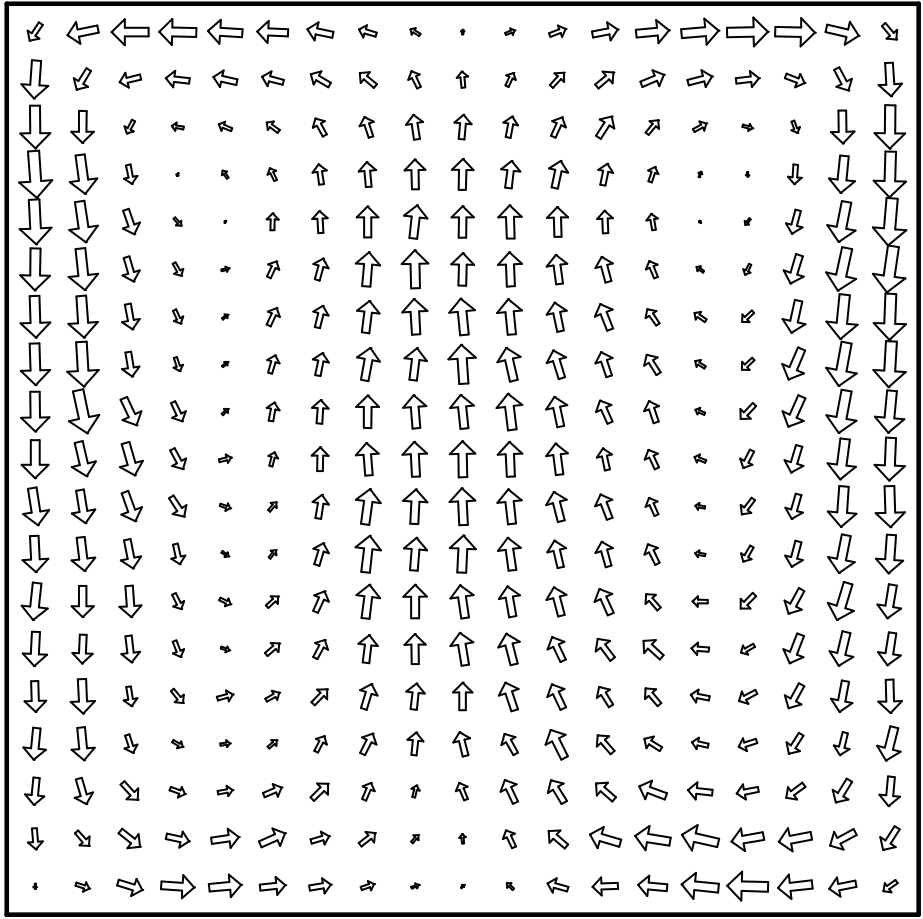
\end{document}